\documentclass[aps,pre,groupedaddress,notitlepage]{revtex4-1}
\usepackage{amsmath,amssymb} % Advanced math typesetting
\usepackage[utf8]{inputenc} % Unicode support (Umlauts etc.)
\usepackage[english]{babel} % Change hyphenation rules
\usepackage{graphicx} % Add pictures to your document
\usepackage{listings} % Source code formatting and highlighting
\usepackage[normalem]{ulem}
\usepackage{soul}
\usepackage{color}

\usepackage{comment}
\usepackage{lineno}
\linenumbers
\makeatletter\AtBeginDocument{\let\@elt\relax}\makeatother

\nolinenumbers
\usepackage{hyperref}
\hypersetup{
    colorlinks=true, %set true if you want colored links
    linktoc=all,     %set to all if you want both sections and subsections linked
    linkcolor=blue,  %choose some color if you want links to stand out
}
\usepackage{xr}
\begin{document}
\title{Onset of criticality in hyper-auxetic polymer networks}

\author{Andrea Ninarello}
\email[]{These authors contributed equally}
\affiliation{CNR Institute of Complex Systems, Uos Sapienza, Piazzale Aldo Moro 2, 00185, Roma, Italy}
\affiliation{Department of Physics, Sapienza University of Rome, Piazzale Aldo Moro 2, 00185 Roma, Italy}

\author{Jos\'e Ruiz-Franco}
\email[]{These authors contributed equally}
\affiliation{CNR Institute of Complex Systems, Uos Sapienza, Piazzale Aldo Moro 2, 00185, Roma, Italy}
\affiliation{Department of Physics, Sapienza University of Rome, Piazzale Aldo Moro 2, 00185 Roma, Italy}

\author{Emanuela Zaccarelli}
\email[Corresponding author:]{emanuela.zaccarelli@cnr.it}
\affiliation{CNR Institute of Complex Systems, Uos Sapienza, Piazzale Aldo Moro 2, 00185, Roma, Italy}
\affiliation{Department of Physics, Sapienza University of Rome, Piazzale Aldo Moro 2, 00185 Roma, Italy}

\date{\today}

\maketitle
{\bf Against common sense, auxetic materials expand or contract perpendicularly when stretched or compressed, respectively, by uniaxial strain, being characterized by a negative Poisson's ratio $\nu$. The amount of deformation in response to the applied force can be at most equal to the imposed one, so that $\nu=-1$ is the lowest bound for the mechanical stability of solids, a condition here defined as ``hyper-auxeticity''. In this work, we numerically show that ultra-low-crosslinked polymer networks under tension display hyper-auxetic behavior at a finite crosslinker concentration. At this point, the nearby mechanical instability triggers the onset of a critical-like transition between two states of different densities. This phenomenon displays similar features as well as important differences with respect to gas-liquid phase separation. Since our model is able to faithfully describe real-world hydrogels, the present results can be readily tested in laboratory experiments, paving the way to explore this unconventional phase behavior.}

\begin{figure}[!h]
\centering
	    \includegraphics[width=0.8\linewidth]{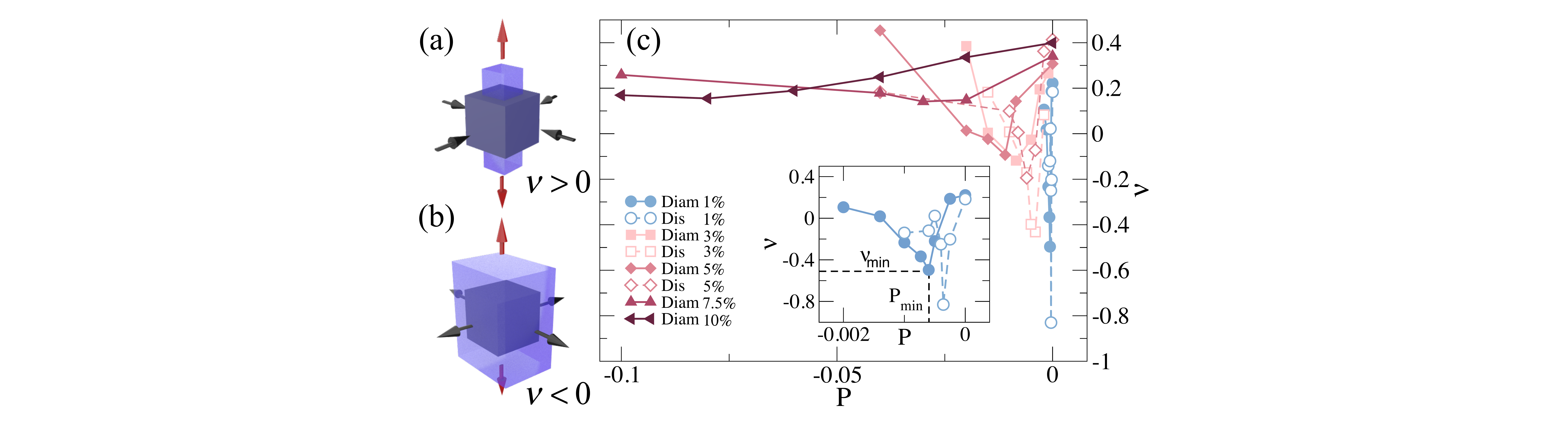}
\centering	    
	    \caption{{\bf Auxetic behavior of ordered and disordered hydrogels.} Illustration of (a) standard ($\nu >0$) versus (b) auxetic ($\nu <0$) behavior. Here, the red arrows indicate the uniaxial deformation (stretching) that is applied on the initial state of the system, represented by the dark cube that is identical in both cases. Following external strain, the system deforms perpendicularly in the directions indicated by the black arrows, leading to two different final states, represented by the light polyhedra;  (c) calculated Poisson's ratio $\nu$ as a function of negative pressure $P$ for different values of crosslinker concentration $c=1,3,5,7.5\%$ both for ordered (Diam, full symbols) and disordered (Dis, empty symbols) networks. Inset: zoom of  $c=1\%$, where the minimum values of pressure and Poisson's ratio, $P_{min}$ and $\nu_{min}$ respectively, are indicated for the Diam-1\% case. Pressure is given in units of $k_B T/\sigma^3$ as described in Methods.
	  \label{fig:nu_min}
	    }
  \end{figure}

The mechanical response of a material subjected to uniaxial strain in the direction orthogonal to the deformation is quantified via the Poisson's ratio $\nu$, defined as the negative ratio between transverse and longitudinal deformation. For the most common three-dimensional materials  $\nu$ is positive, so that these expand (contract) in response to a compressive (extensional) strain. This situation is schematically illustrated in Fig.~\ref{fig:nu_min}(a). On the contrary, auxetic materials are characterized by negative values of $\nu$, meaning that they become thicker perpendicularly to the deformation axis, as shown in Fig.~\ref{fig:nu_min}(b). Auxetic behaviour has been so far reported in a large variety of systems, including foams, polymers, fibers, tendons and crystals~\cite{Lakes1987, Evans1989,evans1991molecular,Gatt2015,Hu2017, Rysaeva2018,greaves2011poisson}.  Recently, a strong research interest has been devoted towards auxetic metamaterials in which the elastic properties can be tailored by geometrical design~\cite{Larsen1997,Theocaris1997,Hanifpour2018} or by pruning methods~\cite{Reid2018}. 
Besides geometrical reasons, a negative $\nu$ can also be obtained by exploiting critical behavior and phase transitions, as in the case of ferroelastic materials in the vicinity of the Curie point~\cite{Dong2010,Kou2016}.

Within linear elasticity theory~\cite{landau1970theory}, the appearance of a negative $\nu$ can be related to a decrease of the bulk modulus $K$ with respect to the shear modulus $G$, namely to an isotropic softening of the material.  A vanishing $K$ echoes the divergence of the isothermal compressibility occurring at a gas-liquid critical point. However, the presence of a finite shear modulus, as found in polymer networks, such as hydrogels, may induce a negative $\nu$. Pioneering evidence of a negative Poisson's ratio has been reported for these systems close to the so-called Volume Phase Transition~\cite{Hirotsu1991, Li1993, Hirotsu1994}: in this case, a variation in temperature changes the affinity of the polymer to the solvent, favoring monomer-monomer aggregation, in full analogy with the gas-liquid critical point, but with the additional constraint of network connectivity. Another thermodynamic parameter that influences the network properties without affecting monomeric interactions is pressure, or tension. Indeed, theoretical works have addressed the occurrence of auxeticity in two-dimensional models of networks under tension~\cite{Wojciechowski1989,Boal1993}.

It is important to notice that these studies have flourished about twenty years ago, but the interest in hydrogel and microgel networks has increased again in the last few years, thanks to advances in chemical and in silico synthesis. In particular, it became recently possible to tune the amount of branching points (crosslinkers) to yield ultra-low-crosslinked networks~\cite{Bachman2015, Virtanen2016,Scotti2019,Scotti2019b}. In parallel, numerical efforts have been able to realize fully-connected, disordered networks with arbitrary density and crosslinker concentrations~\cite{Gnan2017,sorichetti2021effect}. 

In this article, we numerically investigate the elastic properties of polymer networks under tension and demonstrate that auxeticity naturally emerges in the ultra-low-crosslinked limit. Combining stress-strain and equilibrium simulations, we show that low-density polymeric networks exhibit a nonmonotonic behavior of $K$ as well as $\nu$, the latter becoming increasingly negative with reducing crosslinker concentration $c$. This phenomenology is found for both ordered diamond-like and disordered hydrogel realizations, indicating that there is no need of a specific topology to observe auxeticity in polymer networks. Remarkably, we do not find that this behavior continuously evolves down to $c\rightarrow 0$. Rather, it hits a mechanical critical point, that we name hyper-auxetic point, at a finite $c=c^*\sim 0.35\%$ where $\nu=-1$. Owing to the fact that $K$ and $G$ would become negative, this value denotes the lowest limit of mechanical stability, even though this condition is not yet fully understood~\cite{Shan07, greaves2011poisson, Nicolaou2012}. At this point, we detect the onset of a coexistence between two different states: a low density and a high density one. These results call for an analogy with the well-known gas-liquid phase transition in attractive fluids, but with two important differences: (i) the lack of attraction between the monomers due to the  good solvent conditions of the polymer networks and (ii) the presence of critical-like density fluctuations, which do not seem to obey Ising-like statistics within the present numerical resolution. 

\noindent \textbf{Results}\\
\textbf{Auxetic behavior of ordered and disordered hydrogels.} We start by calculating the elastic properties of diamond networks (Diam) for different values of the crosslinker concentration $c$ for negative pressures starting from $P=0$.  Although the diamond network is a simplified model~\cite{rovigatti2019numerical}, it can still describe some relevant phenomena in experiments at a qualitative level~\cite{Keidel2018, Alvarez2019}. For each studied state point, we independently evaluate three moduli:  the bulk modulus $K$ is obtained from equilibrium NPT runs, whereas the Young modulus $Y$ and the Poisson's ratio $\nu$ are estimated from strain-stress simulations, as described in the Methods section. We report $\nu$ in Fig.~\ref{fig:nu_min}(c) for hydrogels with different $c$, while the corresponding $K$ and $Y$ are shown in Fig.~\ref{fig:moduli} of the Supplementary Information. All moduli display a similar behavior: they initially decrease, then reach a minimum at intermediate values of pressures, and finally, increase again for very negative $P$.  The minima become more and more pronounced for lower and lower $c$, remaining visible at all $c$ for $\nu$ and $K$,  while disappearing for $Y$ when $c\gtrsim 3\%$.  Remarkably, we find that the different networks display a positive value of $\nu$ both for $P=0$ and for very large negative pressures, while auxetic behavior is observed  for $c\lesssim 5\%$ in a finite range of tensions, that become smaller and closer to zero pressure as $c$ decreases. We denote the minimum value reached by $\nu$ as $\nu_{min}$ and its corresponding pressure as $P_{min}$ (see inset of Fig.~\ref{fig:nu_min}).

So far we exclusively discussed ordered networks. However, in real-world realizations, hydrogels are intrinsically disordered and frequently made of chains whose length is exponentially distributed~\cite{higgs1988polydisperse,grest1990statistical}. Being able to prepare hydrogels with these features, as described in Methods, we find that disordered networks (Dis) show the same phenomenology as ordered ones when subjected to tensions, with auxeticity also emerging for $c\lesssim 5\%$ (see Fig.~\ref{fig:nu_min}). Interestingly, we observe lower values of $\nu_{min}$ for disordered hydrogels with respect to ordered ones at the same crosslinker concentration, as shown in the inset of Fig.~\ref{fig:nu_min}. In particular, we find $\nu_{min}\approx -0.8$ for the Dis-$1\%$ network. We ascribe this effect to the higher structural heterogeneity characterizing disordered systems, independently of the specific network topology. Indeed, the same qualitative phenomenology is observed for all examined independent realizations (see Fig.~\ref{fig:dis}). Since the preparation procedure is based on the self-assembly of patchy particles~\cite{Gnan2017,sorichetti2021effect}, we cannot easily obtain disordered networks with smaller values of $c$, due to the nearby occurrence of phase separation. Thus, for lower degree of crosslinking, we focus only on diamond networks, having shown in Fig.~\ref{fig:nu_min} that there is no major qualitative effect of the underlying topology on auxeticity, in line with previous results for ordered or partially ordered topologies~\cite{Larsen1997,Theocaris1997,Hanifpour2018,Reid2018}.

\begin{figure}[t]
\centering
	\includegraphics[width=1.0\linewidth]{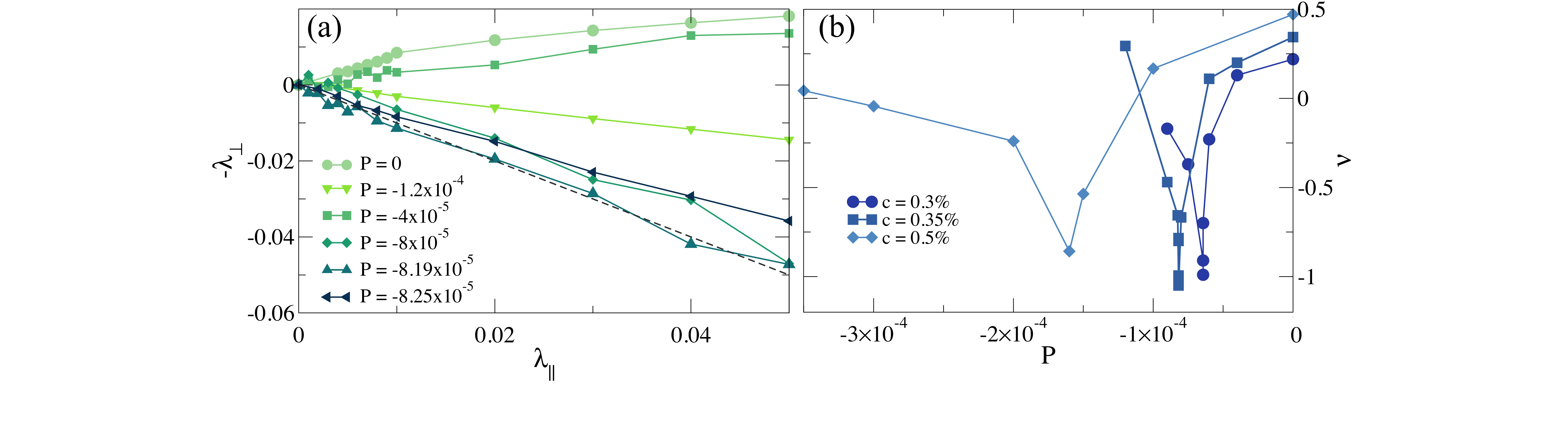}	
	\caption{{\bf Hyper-auxeticity.} (a) Negative transverse strain ($\lambda_\perp$) as a function of longitudinal strain ($\lambda_\parallel$) for $c=0.35\%$ system for different values of  pressure. The dashed straight line has negative unitary coefficient well-approximating the data for $P=P_{min}=-8.19\times 10^{-5}k_BT/\sigma^3$ (hyper-auxetic point); (b) Poisson's ratio $\nu$ as a function of negative $P$ for diamond hydrogels with $c=0.30\%, 0.35\%, 0.5\%$. }
	\label{fig:nu_min_P}
\end{figure}

\textbf{Hyper-auxeticity and mechanical instability.} 
By further decreasing $c$ for Diam-N we still observe a progressive decrease of $\nu_{min}$. To have a better understanding of the mechanical behaviour of the system, we report in Fig.~\ref{fig:nu_min_P}(a) the negative transverse $\left(\lambda_{\bot}\right)$ vs the longitudinal strain $\left(\lambda_{\parallel}\right)$ (see Methods for definition) for $c=0.35\%$ at different values of pressure. We clearly see that, for $P=0$, the slope, that is precisely our numerical estimate of $\nu$, is positive. Then, by progressively reducing $P$,  $\nu$ becomes negative, down to $\nu_{min} \simeq -1$ within numerical uncertainty. At this point, the system has reached the limit of mechanical stability, so that state points with $\nu<-1$ are not allowed. We therefore consider $c^*\simeq 0.35\%$ as the critical fraction of crosslinker at which a mechanical critical point is encountered and define state points with $\nu=-1$ as hyper-auxetic. Interestingly, we find that, upon further increasing tension, the slope starts to increase again, as shown in Fig.~\ref{fig:nu_min_P}(a).  The behavior of $\nu$ vs $P$ is reported for three ultra-low values of $c$ respectively above, at and below $c^*$ in Fig.~\ref{fig:nu_min_P}(b).  These findings indicate that the system reaches a hyper-auxetic condition also for $c < c^*$. This phenomenology thus occurs for ultra-low-crosslinked networks in general, with the system reaching $\nu_{min}=-1$ at a small, finite negative pressure.

\begin{figure}[th]
\centering
	\includegraphics[width=0.7\linewidth]{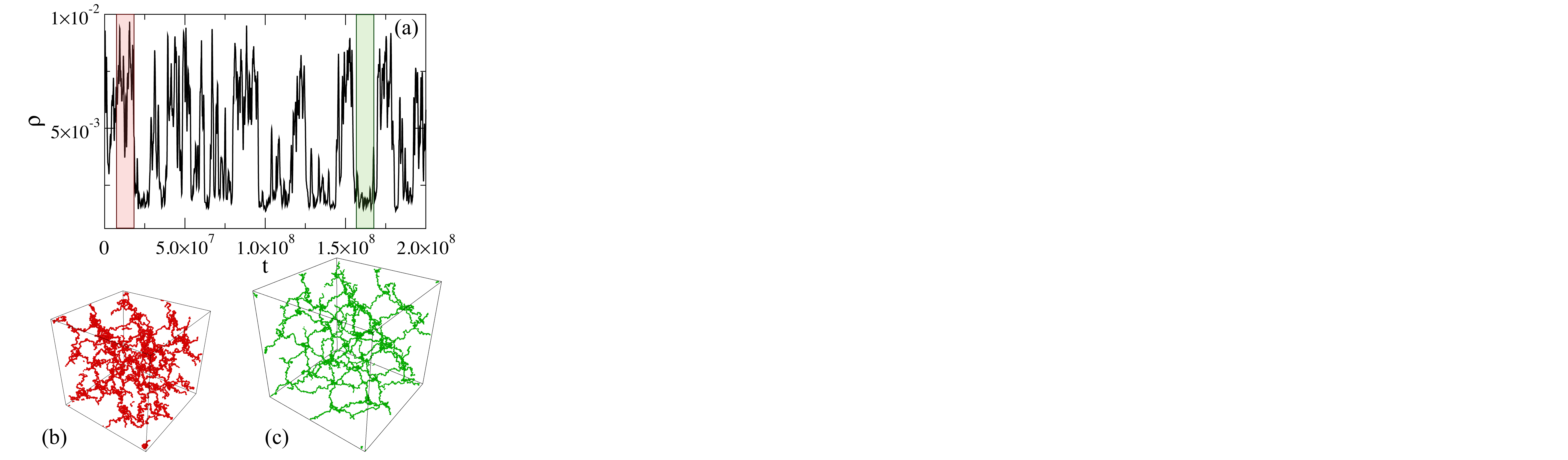}
	\caption{{\bf Density fluctuations and associated elastic properties.} (a) Density fluctuations as a function of time for the Diam - 0.35\% network for $P=P_{min}=-8.19\times 10^{-5}k_BT/\sigma^3$ and corresponding  simulation snapshots of the compressed (b) and expanded states (c) among which the system fluctuates.}
\label{fig:two-states}
\end{figure}

 \begin{figure}[th]
\centering
	\includegraphics[width=1.0\linewidth]{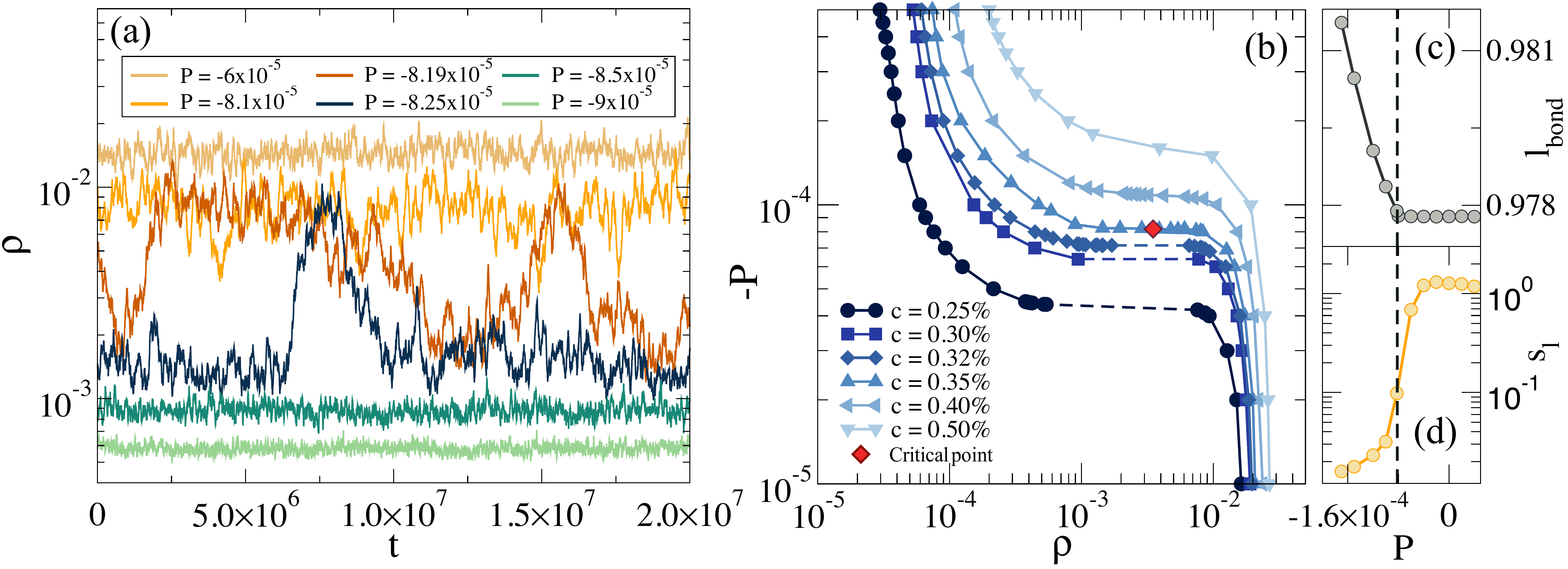}
	\caption{{\bf Phase behaviour and thermodynamics.}
	   (a) Density as a function of time for the Diam - 0.35\% network at different negative pressures; (b)  \textit{Equation of state} of the diamond networks showing negative $P$ as a function of $\rho$ for different values of $c$. The red diamond symbol indicates the approximate location of the onset of mechanical instability and critical-like behavior. To improve visualization, data are shown using the opposite sign for (negative) pressure on a log-log scale; average values of (c) bond length $l_{bond}$ and (d) single-chain entropy $s_l$ as a function of $P$ for the Diam - 0.35\% network. The dashed line indicates the hyper-auxetic point ($P_{min}=-8.19\times 10^{-5}k_BT/\sigma^3$). }
\label{fig:critical}
\end{figure}

{\bf Critical-like nature of the transition.} 
The behaviour discussed so far close to hyper-auxeticity may echo what happens close to a thermodynamic second-order phase transition, such as the gas-liquid one, where a homogeneous system becomes thermodynamically unstable due to the divergence of the isothermal compressibility. To avoid this, the system thus separates into two phases characterized by a different density. It is now interesting to investigate by which mechanism ultra-low-crosslinked hydrogels deal with the presence of the mechanical instability and which similarities or differences with respect to the gas-liquid scenario occur. To this aim, we report the behavior of the density fluctuations with respect to time for the Diam - 0.35\% system at $P=P_{min}$ in Fig.~\ref{fig:two-states}(a), detecting the onset of critical-like fluctuations. It is evident that, close to the mechanical instability, the system fluctuates between two states, an expanded and a compressed one, respectively illustrated in the corresponding snapshots of Fig.~\ref{fig:two-states}(b,c). We thus separately calculate the elastic moduli of the two states and find that the bulk modulus is much smaller in the expanded case ($K_{\textrm{EXPANDED}}\sim 1.4\times 10^{-6} k_BT/\sigma^{3}$) as compared to the compressed one ($K_{\textrm{COMPRESSED}}\sim 5.0\times 10^{-6}  k_BT/\sigma^{3}$). A similar behavior was also detected for the Young modulus with the compressed state having it significantly larger than the expanded one. However, the Poisson's ratio does not change so much, being $\nu \simeq -1$ for the expanded state and $-0.84$ for the compressed one, as shown in Fig.~\ref{fig:nu_exp_compr}. These data also confirm the absence of relevant anisotropic effect or preferential directions within our system and suggest that a hyper-auxetic behavior is found in both states. Hence, the mechanism of a density jump is the one allowing the system to avoid the mechanical instability in full analogy with gas-liquid phase separation.  

Such an analogy is clearly evident when looking at the density fluctuations as a function of time for different pressures around $P_{min}$ for the Diam - 0.35\% network, that are shown in Fig.~\ref{fig:critical}(a). We now examine what happens as a function of $c$ and plot the behavior of the (negative) pressure against density in Fig.~\ref{fig:critical}(b) for all studied diamond networks. It is important to note that, in the present model, the crosslinker concentration plays the role of temperature in phase-separating fluids: as $c$ becomes smaller, $P$ becomes progressively flatter and a critical-like point is observed for these putative equation of states, indicated by a red diamond in the figure, which depend on a geometric rather than a thermodynamic control variable. For $c\leq c^*$, a clear discontinuity in density is observed, signaling a first-order-like transition between the two states. To get better microscopic insights of this behavior, we report the average bond length $l_{bond}$ as a function of $P$ in Fig.~\ref{fig:critical}(c). From a geometrical perspective, lowering the pressure towards more negative values has the effect of stretching the chains, as demonstrated by the growth of $l_{bond}$. However, it is important to note that this behavior arises only when $P\approx P_{min}$, i.e. when $\nu$ starts to increase again beyond its minimum value. Since it is well-known that entropy plays a fundamental role for phase behaviour of polymer systems~\cite{rubinstein2003polymer}, we also quantify the effect of entropy by calculating the average single-chain entropy $s_l$ within a Langevin-approximation (see Methods and Ref.~\cite{sorichetti2021effect}). Relying on this controlled approximation, we detect a decrease of $s_l$ by roughly one order of magnitude going from the compressed to the expanded state, as shown in Figs.~\ref{fig:critical}(d).  The entropy further shows critical-like fluctuations close to $P_{min}$, as shown in Fig.~\ref{fig:entropy}. 

On the other hand, when we monitor the average total potential energy  $e_t$, we find no remarkable change with pressure and, importantly, no critical fluctuations, as shown in Fig.~\ref{fig:ene_rho}(a). We thus focus on the non-bonded potential energy $e_{nb}$, pertinent only to non-bonded particles, shown in Fig.~\ref{fig:ene_rho}(b), which instead manifests critical-like fluctuations. This correspondence is confirmed by the scatter plots, reported in Fig.~\ref{fig:ene_rho}(c) and (d), of total energy and non-bonded energy with density, respectively. Clearly, correlation is completely absent between $e_t$ and $\rho$, while $e_{nb}$ is correlated with density, similarly to the single chain entropy as discussed in the SI.

\begin{figure}[th]
\includegraphics[width=1.0\linewidth]{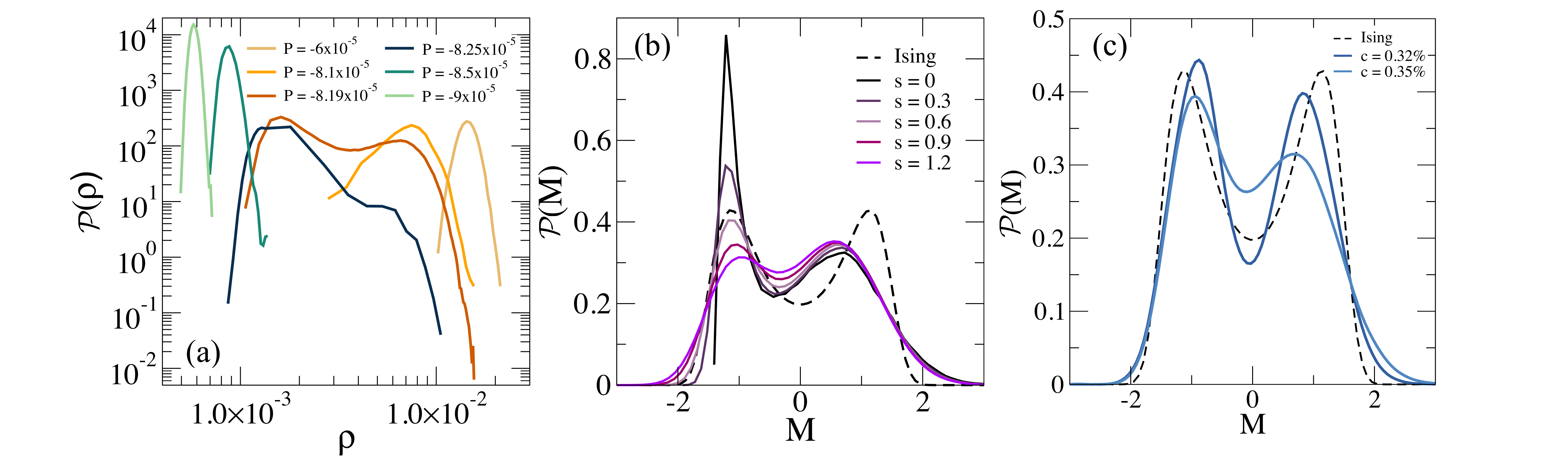}
\caption{\label{fig:prho} 
{\bf Density distributions and comparison with Ising statistics.}
(a) Density distribution $\mathcal{P}(\rho)$ for different negative pressures for the Diam - 0.35\% network; (b) distribution $\mathcal{P}(M)$ of the order parameter $M=\rho+s e_{nb}$, where $s$ is the mixing parameter and $e_{nb}$ is the non-bonded particles potential energy,  at $P_{min}=-8.19\times 10^{-5}k_BT/\sigma^3$ for different values of $s$. The dashed line is the reference Ising curve~\cite{Wilding1992}; (c) optimized $\mathcal{P}(M)$ obtained from histogram reweighting for Diam - 0.32\% and 0.35\% networks. The curves are reported for the state points $P=-6.8898\times 10^{-5}k_BT/\sigma^3, s=2.59$ and $P=-8.9135\times 10^{-5}k_BT/\sigma^3, s=1.25$, respectively, representing distributions that are found to be closest with respect to the Ising one through single histogram reweighting in pressure (see Methods for details).	 
 }
\end{figure}

{\bf Density distributions and comparison with Ising statistics.} We now examine in more detail the nature of the density fluctuations and report the distribution of the density $\mathcal{P}(\rho)$ in Fig.~\ref{fig:prho}(a) for different negative pressures close to the onset of the mechanical instability for the Diam - 0.35\% network.  We notice that, close to the transition, the system displays a bimodal distribution, that is highly asymmetric. This is even more evident from the fact that data are reported on a log-log scale to improve visualization. In particular, we find the presence of a broad high-density peak and a narrow low-density one peak for low ones. 
Aiming to build a correspondence with thermodynamic phase separation, we next calculate the order parameter $M$, equivalent to the one used to describe the gas-liquid transition, that is composed by density and energy fluctuations~\cite{Wilding1992, Sciortino2020}. Since we found that for the present system the total potential energy is not correlated with density, we define $M=\rho+se_{nb}$, where we only consider the non-bonded potential energy and $s$ is a mixing parameter~\cite{Wilding1992}. In Fig.~\ref{fig:prho}(b), we plot the distribution of the order parameter $\mathcal{P}(M)$ with zero average and unit variance for $P=P_{min}$ and different values of $s$. We find that the presence of the mixing term is able to reduce the asymmetry of the original $\mathcal{P}(\rho)$, but still not completely. In particular, the heights of the two peaks become comparable for $s=0.9$, but the difference in their variance is retained at all studied $s$. 

In order to compare with the expected 3D Ising universal distribution, we apply the single histogram reweighting technique, as discussed in the Methods. This is shown in Fig.~\ref{fig:prho}(c) for state points close to the mechanical instability at two different values of $c$. The resulting $\mathcal{P}(M)$ for ultra-low-crosslinked hydrogels are characterized by a slightly asymmetric shape, not perfectly centered in zero with rather broad peaks. We thus find a qualitative disagreement with the Ising behavior, independently of $c$, that might be due either to an intrinsic difference of the network system with respect to associating molecules or to insufficient sampling. Assuming true the first case, we may speculate that the presence of the network connectivity may bias the way in which the density fluctuates as compared to unbound particles. Alternatively, the deviation may be attributed to the fact that, in the present simulations, $c$ cannot be varied in a continuous way, as normally done with temperature close to the gas-liquid critical point, thus hindering a proper exploration of the critical properties of the transition. However, we note that we found deviations from the Ising expectations for all examined $c$ values, moving either below or above $c^*$, finding no systematic improvement. Future work will be needed to properly address this issue.

\noindent\textbf{Discussion}\\
The present results, obtained by means of extensive simulations to calculate the elastic properties of ultra-low crosslinked polymer networks at negative pressures, report the emergence of auxetic behavior for $c \lesssim 5\%$ independently of the specific geometry of the network. It is important to note that, when we started our investigations at not-too-small values of $c$ and detected the onset of auxeticity, we expected to find a continuous behavior of the system until approaching the limit of stability ($\nu=-1$) at $P=0$. This was indeed suggested by looking at the $c$-dependence of $P_{min}$ and $\nu_{min}$, that is reported in Fig.~\ref{Pmin-numin}: while $\nu_{min}$ decreases logarithmically, $P_{min}$ is found to obey a quadratic power-law behavior for all examined networks which spans a range of more than three decades in $c$.  Instead, surprisingly, we found that the limit of the mechanical stability of the system, and the associated hyper-auxetic behavior ($\nu=-1$), occurs at a finite crosslinker concentration, $c^*\simeq 0.35\%$, even for a regular network such as the diamond one. Hence, for even smaller values of $c$, the network, being unstable, undergoes a transition between two states, a compressed and an expanded one. Such a transition is also accompanied by large density fluctuations reminiscent of critical ones in gas-liquid phase separation, as shown in Fig.~\ref{fig:critical}(a).

We then examined in more detail the nature of this phase transition, that is clearly distinct from the widely investigated Volume Phase Transition (VPT) of thermoresponsive hydrogels. Indeed, the latter occurs due to the change of the underlying polymer-solvent interactions at a characteristic temperatures. Previous studies on the VPT of hydrogels have already focused on the associated critical properties, that were tentatively attributed to the Ising university class~\cite{Li1989, Onuki1993,Seiffert2014}.  Instead, the present work focuses on networks in good solvent conditions, where the monomer affinity to the solvent does not change and the underlying interactions are always dominated by excluded volume. Thus, the phase transition observed in the present work is the consequence of changing the network connectivity down to very low $c$, which generates a non-trivial interplay with steric interactions under a small tension. To this aim, it is instructive to focus on the values of the pressure at which $\nu$ reaches its minimum for $c^*$, i.e., $P_{min}\sim10^{-4} k_BT/\sigma^3$, while the system volume $V$ fluctuates around $\sim 10^6-10^7\sigma^3$. This implies that the product $PV$ is roughly comparable to the scale of the non-bonded particle energy $\sim 0.02N k_BT$ and about $3-4$ orders of magnitude smaller than the total energy scale of the system $\sim 20 N k_BT$ (see Fig.~\ref{fig:ene_rho}), where $N\approx 10^4$ is the total number of monomers in our simulations. These considerations confirm that the present phase transition is dominated by fluctuations of non-bonded energy and of entropy, with a negligible contribution of the total energy. The important role of (infinite) connectivity and the different interactions with respect to a standard attractive system (e.g. a Lennard-Jones fluid) may thus be the reasons why the critical-like fluctuations of the present ultra-low-crosslinked hydrogels are not found to obey Ising universality class. Further numerical and theoretical work on this issue will be needed in the future. While the former should aim, in particular, to probe the critical fluctuations in a more extensive time and length window as well as to vary $c$ in a more continuous fashion (e.g. by developing appropriate crosslinker insertion/deletion methods), the latter should be devoted to provide an additional description of the mechanical instability, taking into account the connectivity, similarly to what discussed for the Volume Phase Transition~\cite{Fernandez2019}. 

Finally, it is important to note that these observations are relevant for ultra-low-crosslinked polymer networks, that are nowadays within experimental reach~\cite{Gao2003,Bachman2015,scotti2020flow}. Indeed, for example, Poly(N-isopropylacrylamide) (PNIPAm) microgels are synthesized even in the absence of crosslinkers, taking advantage of (rare) self-association of NIPAM monomers, which gives rise to an effective $c$ in the system that is close to zero~\cite{Bachman2015,virtanen2016persulfate}. It should thus not be difficult to realize this also for hydrogels. Given that experimental realizations are necessarily disordered, our numerical predictions suggest that disordered networks should display a slightly larger value of $c^*$, not too far from 1\% (see Fig.~\ref{fig:nu_min}), with respect to the diamond case, which would actually favour the experimental observation of this mechanical critical point and a through exploration of its vicinity both from above and from below $c^*$. Notably, our results are based on hydrogel simulations, but it would be interesting to apply our analysis also to ultra-low-crosslinked microgels. In this respect, the microfluidic approach to microgels synthesis appears to be particularly promising, because it allows to prepare microgels of sizes of the order of $100\mu$m~\cite{Seiffert2012,Voudouris2013}. Furthermore, a specific method to calculate their elastic properties, known as capillary micromechanics, was already established, making these ideal model systems to test our numerical predictions. We thus hope that the present results will stimulate novel experimental activity on ultra-low-crosslinked polymeric materials, aiming to verify their peculiar hyper-auxetic behavior and the occurrence of such an unusual phase transition, where mechanical and thermodynamic instabilities appear to be strongly intertwined, opening up a new research direction in statistical and soft matter physics.

\noindent {\bf Methods}\\
{\bf Model.} We perform Molecular Dynamics simulations of polymer networks made of monomers interacting via the Kremer-Grest potential. Excluded volume for all particles are given by the Weeks-Chandler-Andersen potential:~\cite{weeks1971role}
\begin{equation}
\label{eq:WCA}
V_{WCA}\left( r \right )=\left\{\begin{matrix}
 4\epsilon\left[\left(\frac{\sigma}{r} \right )^{12}-\left(\frac{\sigma}{r} \right )^{6} \right ]+\epsilon &\ \ \  if \ \ \  r \leq 2^{1/6}\sigma\\ 
0 &\ \ \  if \ \ \  r > 2^{1/6}\sigma
\end{matrix}\right.
\end{equation}
where $\sigma$ is the monomer diameter, which sets the unit of length, and $\epsilon$ controls the energy scale. Defining $m$ as the mass of the particles, the unit time of our simulations is defined as $\tau=\sqrt{m\sigma^{2}/\epsilon}$. Chemical bonds between connected monomers are modeled by a FENE potential $V_{FENE}\left(r\right)$~\cite{kremer1990dynamics}:

\begin{equation}
\label{eq:FENE}
V_{FENE}\left( r \right )=-\epsilon k_{F}R^{2}_{0}\ln\left[1-\left(\frac{r}{R_{0}\sigma} \right ) \right ] \ \ \  if  \ \ \  r < R_{0}\sigma 
\end{equation}

where $k_{F}=15$ is the spring constant and $R_{0}=1.5$ is the maximum extension of the bond. We consider both ordered (diamond-like) and disordered topologies. In the former case, we prepare systems made up of $8$ unit cells, each containing $8$ crosslinkers, that are placed on the lattice atom positions and are connected through chains of equal length~\cite{claudio2009comparison, jha2011study}. For each crosslinker concentration $c$, the network is composed of $N=64/c$ monomers forming chains of equal length $l=(1-c)/(2c)$,  thus by varying $l$ we change $c$ in a controlled way. To produce disordered networks, we use the method recently developed in Ref.~\cite{Gnan2017,Ninarello2019}, which exploits the self-assembly of binary mixtures of patchy particles with valence $f=2$ (monomers) and $f=4$ (crosslinkers). We let the system equilibrate at low enough temperature ($T=0.03$) through the oxDNA simulation package~\cite{oxDNA} until 99.9\% of the bonds are formed by exploiting a recently devised swap algorithm~\cite{Sciortino2017}. Then, we select the largest cluster from which we remove dangling ends and replace patchy interactions with the bead-spring ones (Eqs.~\ref{eq:WCA} and~\ref{eq:FENE}). We obtain systems with final crosslinker concentrations $c\sim 1,3,5 \%$ with deviation from the nominal values smaller than $5\%$. For $c\sim 1\%$ we consider three independent network realizations to assess the dependence of results on the specific topology. 

For both ordered and disordered networks we perform NPT simulations using LAMMPS simulation package~\cite{plimpton1995fast} with a Nos\'e - Hoover thermostat and barostat. Temperature is set to 1.0 throughout the manuscript and is measured in units of energy, i.e. fixing also $k_B=1$, where $k_{B}$ is the Boltzmann constant. We thus perform simulations at different (negative) pressures employing a timestep $\delta t = 0.003\tau$.

{\bf Calculation of elastic moduli. } We perform two kinds of simulations: (i) via equilibrium simulations in which the box is allowed to fluctuate anisotropically we calculate the bulk modulus $K$ from volume fluctuations as
$ K=k_BT \frac{\langle V\rangle}{\langle V^2 \rangle - \langle V \rangle ^2}$; (ii) via stress-strain simulations we simultaneously calculate $Y$ and $\nu$. In particular, we first apply a longitudinal extensional strain $\lambda_{\parallel}=\left(L_{\parallel}-L_{\parallel}^{0}\right)/L_{\parallel}^{0}$, where $L_{\parallel}^{0}$ and $L_{\parallel}$ are the initial and final box lengths along the deformation axis, respectively. The range of deformation values encompasses $\lambda_{\parallel}\in\left[0, 0.3\right]$, at which the response is in the linear regime, and we use a fixed strain rate $\dot{\lambda}=0.01\tau^{-1}$. The box is allowed to fluctuate transversally to the deformation in order to guarantee a constant average $P$. Then, once the system acquires the desired strain, the stress $\sigma_{\parallel}$ along the deformation axis is calculated from the virial stress tensor and averaged over $10^{6}\tau$, yielding $Y=\sigma_{\parallel}/ \lambda_{\parallel}$. The Poisson's ratio is instead extracted from transversal fluctuations through the relation $\nu=-\partial \lambda_{\bot}/\partial \lambda_{\parallel}$, where $\lambda_{\bot}\equiv\left(\lambda_{2}+\lambda_{3}\right)/2$ and $\lambda_2$, $\lambda_3$ are the components of the strain orthogonal to the deformation axis. For each network and each state point, results for $Y$ and $\nu$ are averaged over 20 independent deformations in which the same configuration is deformed with different initial velocities taken from the Maxwell-Boltzmann distribution. This procedure is repeated for each configuration by deforming the network over all three directions independently, which are then averaged in order to improve the statistics of the results. Results for $K$ and $Y$ are given in units of $k_BT/\sigma^3$.

{\bf Langevin approximation for single chain entropy.}  In general, the entropy of a chain with $n$ bonds of length $b$ can be written as~\cite{flory1976statistical}
\begin{equation}
s\left(n, r\right)=k_{B}logW_{n}\left(r\right)+A_{n}\,,
\end{equation} 
\noindent where $\mathbf{r}=\left(r_{x}, r_{y}, r_{z}\right)$ is the end-to-end vector of the chain, $W_{n}\left(r\right)$ is the end-to-end probability density and $A_{n}$ is a temperature-dependent parameter that can be set to zero. In the limit of systems that are submitted to a very large deformation or in a dilute regime, i.e., for $r\sim nb$, the end-to-end probability reads as~\cite{treloar1975physics}  
\begin{equation}
W_{n}\left(r\right)\sim \exp\left[-\frac{r}{n}\mathcal{L}^{-1}\left(r/nb\right)\right]\left[\frac{\mathcal{L}^{-1}\left(r/nb\right)}{\sinh\mathcal{L}^{-1}\left(r/nb\right)}\right]^{-n}\,,
\label{eq:ProbLangevin}
\end{equation}
\noindent where $\beta=1/k_{B}T$ and $\mathcal{L}^{-1}\left(r/nb\right)$ is the inverse Langevin function, defined as $\mathcal{L}\left(r/nb\right)=\coth\left(r/nb\right)-nb/r$~\cite{jedynak2015approximation}. Thus, the entropy of a single chain can be expressed as:
\begin{equation}
s_{l} \left(n,r\right)= k_B\left[-\frac{r}{b}\mathcal{L}^{-1} \left(r/nb\right)\right] \ln\left[\left(\frac{\mathcal{L}^{-1}\left(r/nb\right)}{\sinh\mathcal{L}^{-1}\left(r/nb\right)}\right)^{-n} \right] + A_n\,,
\label{eq:end-to-end}
\end{equation}
\noindent where, in our case, $b$ is the minimum value of the FENE interaction potential. We use this equation to calculate the single chain entropy of each chain in the network and then average over all chains to obtain the average chain entropy $s_l$ that is reported in Fig.~\ref{fig:critical}(d) and in the SI.

\textbf{Ising Universality class. }
A universality class groups phenomena arising in different physical systems, that, although being described by diverse microscopic models, exhibit an asymptotic large-scale limit that is characterized by the same invariant critical exponents~\cite{binney1992theory}. In particular, the universality class defined by the three-dimensional Ising model describes the phenomenology of second-order phase transitions as diverse as the ferromagnetic Curie point or the gas-liquid criticality. We thus attempt to compare the critical behavior of ultra-low-crosslinked networks to the expected behavior of a system belonging to the 3D Ising universality class, relying on the asymptotic expression of the probability distribution of the order parameter $\mathcal{P}_{Ising}\left(M\right)$ that can be conveniently approximated as~\cite{tsypin2000probability}:
\begin{equation}
\mathcal{P}_{Ising}\left(M\right)\propto \exp\left[-\left(\gamma M^{2}-1 \right)^{2} \left(a\gamma M^{2} + c \right) \right]\,,
\label{eq:ising}
\end{equation}
\noindent where $a=0.158$, $c=0.776$, and $\gamma$ is adjusted to provide unit variance to the distribution. 

\textbf{Histogram reweighting technique. } We  employ histogram reweighting in order to better identify the putative critical point of our phase transition, since this technique provides a powerful tool to reconstruct the probability distribution of a given observable at a state point $\mathcal{P}\left(P',c'\right)$ from equilibrium distributions of close enough state points, as long as thermodynamical control variables vary continuously. This is not the case for our system, for which $c$ assumes discrete values in the ordered system and cannot be finely controlled in the disordered system, as a result of the self-assembly procedure of the network. We are thus forced to perform histogram reweighting only in $P$ and, to this aim, we perform numerous, long-time $NPT$ simulations at each $c$ around $P_{min}$. During the simulations, we record the behavior of the density $\rho$ and of the non-bonded particle energy $e_{nb}$ as a function of time. Then, our single histogram reweighting method, reported in Fig.~\ref{fig:prho}(c), relies on the following expression:
\begin{equation}
\mathcal{P}\left(V,e_{nb}; P'\right)=\mathcal{P}\left(V,e_{nb}; P\right)\exp\left[\left(P-P'\right)V \right] \,.
\end{equation}
\noindent Thus, for each $P'$, we  calculate the histogram reweighting factor $\exp\left[\left(P-P'\right)V \right]$. We use this expression to obtain the distribution of the order parameter $M=\rho + s e_{nb}$, as discussed in the main text. This is calculated for all the values of $\rho$ and $e_{nb}$ in $\mathcal{P}\left(V,e_{nb}; P'\right)$ by varying the mixing parameter $s$. The set of $M$ values is then rescaled to have a zero mean and a unit variance and compiled into the histogram $\mathcal{P}\left(M\right)$. This last is finally scaled onto the Ising curve for each given $c$ by minimizing the mean squared error between the two distribution by varying $P'$.
\\
\\
\centerline{ \textbf{Acknowledgements} }\\
We thank F. Goio Castro, L. Rovigatti and F. Sciortino for useful discussions. We acknowledge support from the European Research Council (ERC Consolidator Grant 681597, MIMIC), from the European Union's Horizon 2020 research and innovation programme (Grant 731019, EUSMI) and from Sapienza University of Rome through the SAPIExcellence program. The authors gratefully acknowledge the computing time granted by EUSMI on the supercomputer JURECA at the J\"ulich Supercomputing Centre (JSC).

\clearpage
     
\section*{Supplementary materials}
\setcounter{equation}{0}
\setcounter{figure}{0}
\setcounter{table}{0}
\setcounter{section}{0}

\renewcommand{\theequation}{S\arabic{equation}}
\renewcommand{\thefigure}{S\arabic{figure}}
\renewcommand{\thetable}{S\arabic{table}}
\renewcommand{\thesection}{S\Roman{section}}

\maketitle
\section{Elastic moduli at ultra-low crosslinking: Comparison with linear elasticity theory}
\label{sec:s1}

\begin{figure}[th]
\includegraphics[width=0.8\linewidth]{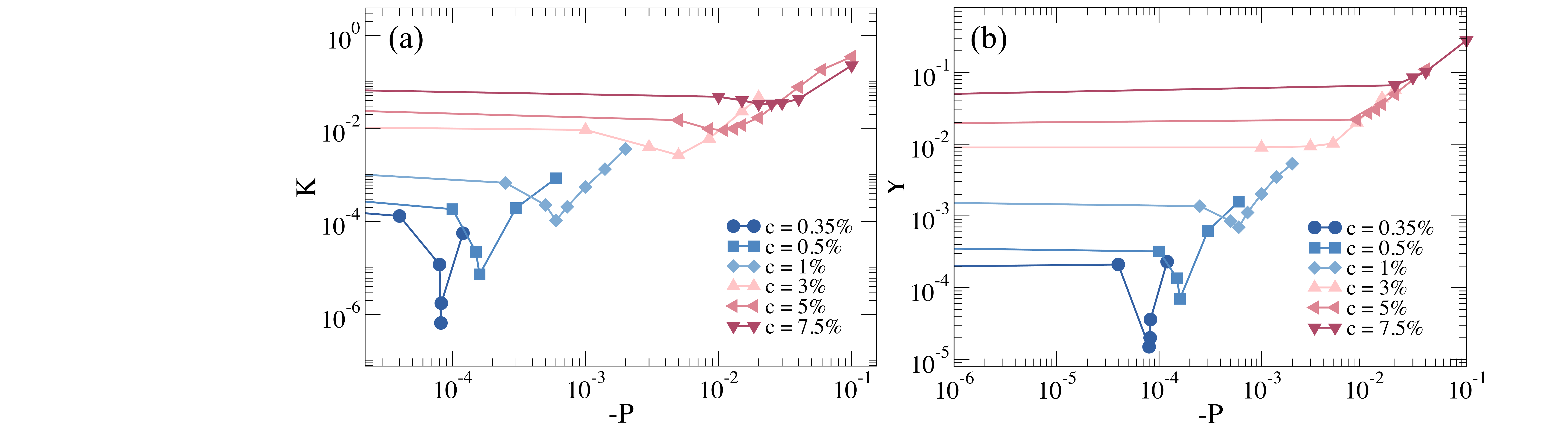}
\caption{\label{fig:moduli} Bulk modulus K (a) and Young modulus (b) as a function of pressure for diamond networks with different crosslinker concentrations. Both moduli are given units of $k_B T/\sigma^3$ (see Methods in the main text).}
\end{figure}

In Fig.~\ref{fig:moduli} we report the bulk $K$ and  Young $Y$ moduli as a function of pressure $P$ for for a diamond topology, with different crosslinker concentrations $c$. Both moduli at low $c$ display a minimum, which becomes more pronounced upon further  lowering $c$, at a characteristic (negative) value of $P$. Such values appear to be close but not exactly the same,  within the current numerical resolution, 
 as $P_{min}$, where the Poisson's ratio has a minimum, as discussed in the main text. 
 
\section{Dependence on network topology for disordered realizations}
\label{sec:s2}

\begin{figure}[ht]
\includegraphics[width=0.8\linewidth]{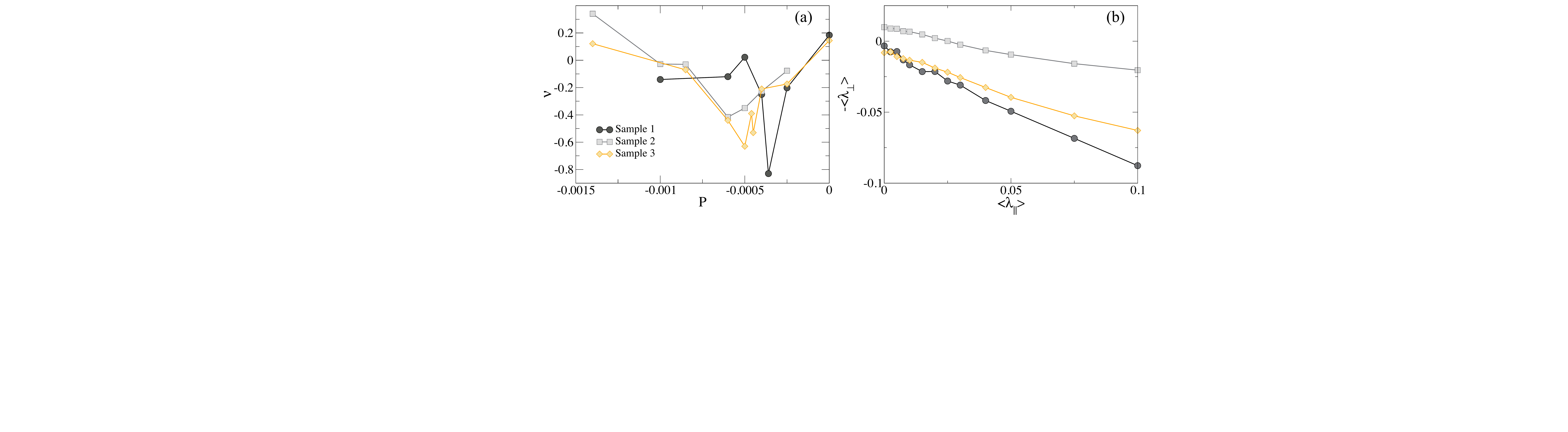}
\caption{\label{fig:dis} 
Stress-strain simulation results for the Dis - 1\% network for three different topologies: (a) Poisson's ratio versus negative pressure; (b) negative transverse strain as a function of parallel strain for each topology at their respective $P_{min}= -3.6 \times 10^{-4}, -5 \times 10^{-4}, -6 \times 10^{-4} k_B T/\sigma^3$ for topology $1,2$ and $3$, respectively. }
\end{figure}
In Fig.~\ref{fig:dis} we report results for the Poisson's ratio for different realizations of the Dis - 1\% network, to show that we find 
consistent behaviour independent of the specific topology. In particular, Fig.~\ref{fig:dis}(a) shows $\nu$ as a function of pressure for the three studied realizations,  displaying a  minimum in all cases, although taking place at different values of $P_{min}$. The transverse versus longitudinal strain for each $P_{min}$ shows a similar behaviour for the different topologies, as reported in Fig.~\ref{fig:dis}(b). 

\section{Elastic properties for compressed and expanded (coexisting) states at the hyper-auxetic point}
\label{sec:s3}

\begin{figure}[h]
\includegraphics[width=0.8\linewidth]{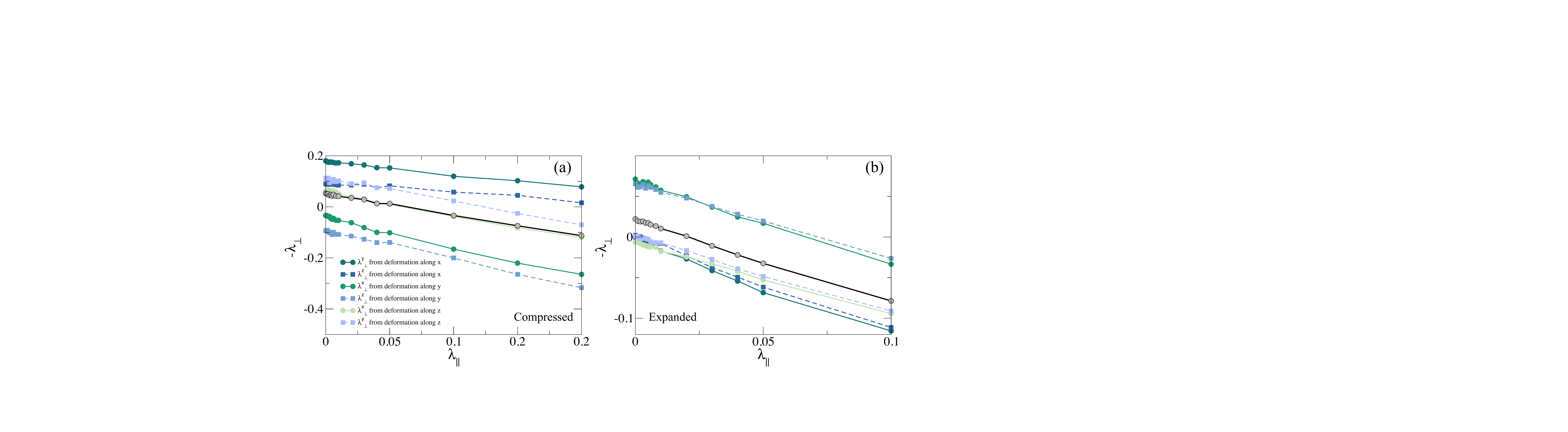}
\caption{\label{fig:nu_exp_compr}  Results for the Diam - 0.35\% network at $P_{min}=8.19\times10^{-5} k_B T/\sigma^3$: negative transverse strain obtained for the three axial directions as a function of parallel strain $\lambda_{\parallel}$ both for the compressed (a) and the expanded (b) state.
}
\end{figure}

Here we report results for the Poisson's ratio of the two coexisting compressed and elastic states for the Diam - 0.35\% network at $P_{min}=8.19\times10^{-5}k_B T/\sigma^3$, as discussed in Fig. 3 of the main text. The calculated $\lambda_\bot$ vs $\lambda_\parallel$ for the compressed and expanded states are reported  in Figs.~\ref{fig:nu_exp_compr}(a),(b), respectively, for all examined deformation directions. From these data, it is evident that the Poisson's ratio does not change much between the two states, being $\nu \simeq -1$ for the expanded state and $-0.84$ for the compressed one.  These data suggest that a hyper-auxetic behaviour is found for both states, although the bulk and the Young moduli are found to be significantly different from each other, as discussed in the main text.

\section{Critical-like behaviour: entropy and energy}

\begin{figure}[th]
\includegraphics[width=0.8\linewidth]{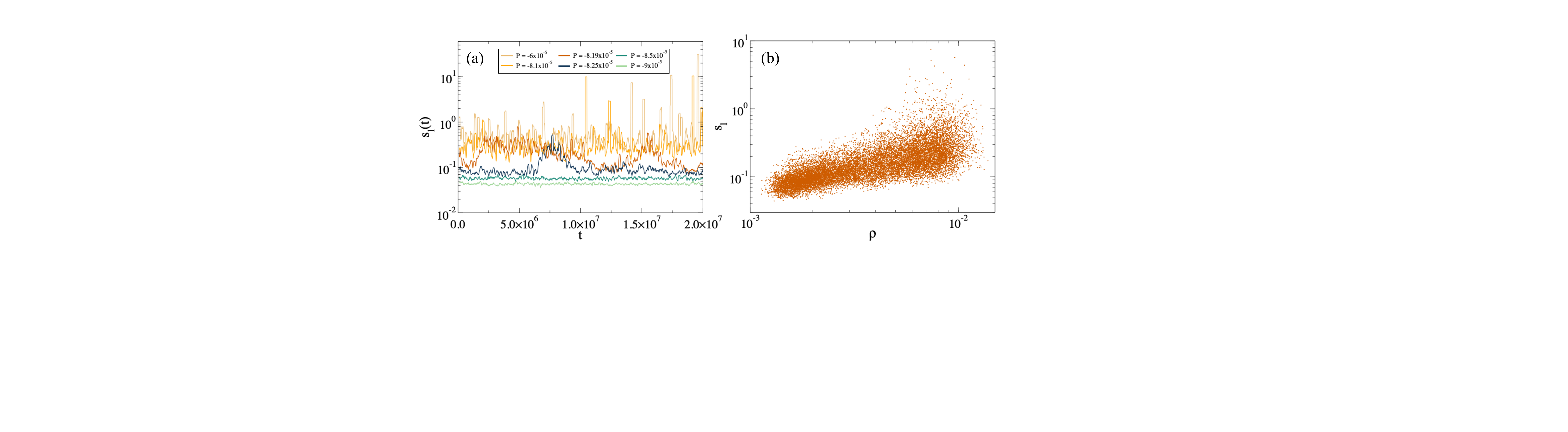}
\caption{\label{fig:entropy}
(a) Single chain entropy $s_l$ as a function of time for the Diam - 0.35\% at different negative pressures;
(b) scatter plot of the same quantity with respect to density at $P_{min}=-8.19\times 10^{-5} k_B T/\sigma^3$.}
\end{figure}

As discussed in the main text, entropy plays an important role in the transition. In polymeric systems, the presence of crosslinkers and entanglements induces a constraint on the configurations that can be explored by the polymer chains. Typically, the chain end-to-end distance $r$ can be used to describe chains fluctuations, whose probability distribution has theoretically been approximated by a Gaussian distribution~\cite{mark2007physical,rubinstein2003polymer}. In this way, the entropy of the network is computed as the sum of the contributions from every chain. While this approach was shown to be useful in dense systems, in dilute regimes it fails due to the presence of short chains whose $r$ does not behave in a Gaussian way. Furthermore, by stretching the network, deviations from the gaussian behavior will be significant~\cite{sorichetti2021effect}. To avoid these issues, here we calculate single chain entropy assuming that the end-to-end distance follows the Langevin approximation~\cite{rubinstein2003polymer}, which for a single chain is given by Eq. 5 of the main text. This approach has been shown to work relatively well in the case of phantom networks up to end-to-end distances approaching the contour length $nb$~\cite{sorichetti2021effect}. Thus, we define the average single chain entropy $s_l$ as the average entropy of all the chains within the network which is reported in Fig.~\ref{fig:entropy}(a) for the Diam - 0.35\% network at different negative pressures. Interestingly, we observe critical-like jumps between a high entropy regime corresponding to high density and low entropy states corresponding to low density, which happen simultaneously with those occurring for density (see Fig.~\ref{fig:critical}(a) in the main text). This is confirmed by the scatter plot of entropy and density at $P_{min}$ in Fig.~\ref{fig:entropy}(b).
 
 \begin{figure}[h]
\includegraphics[width=0.8\linewidth]{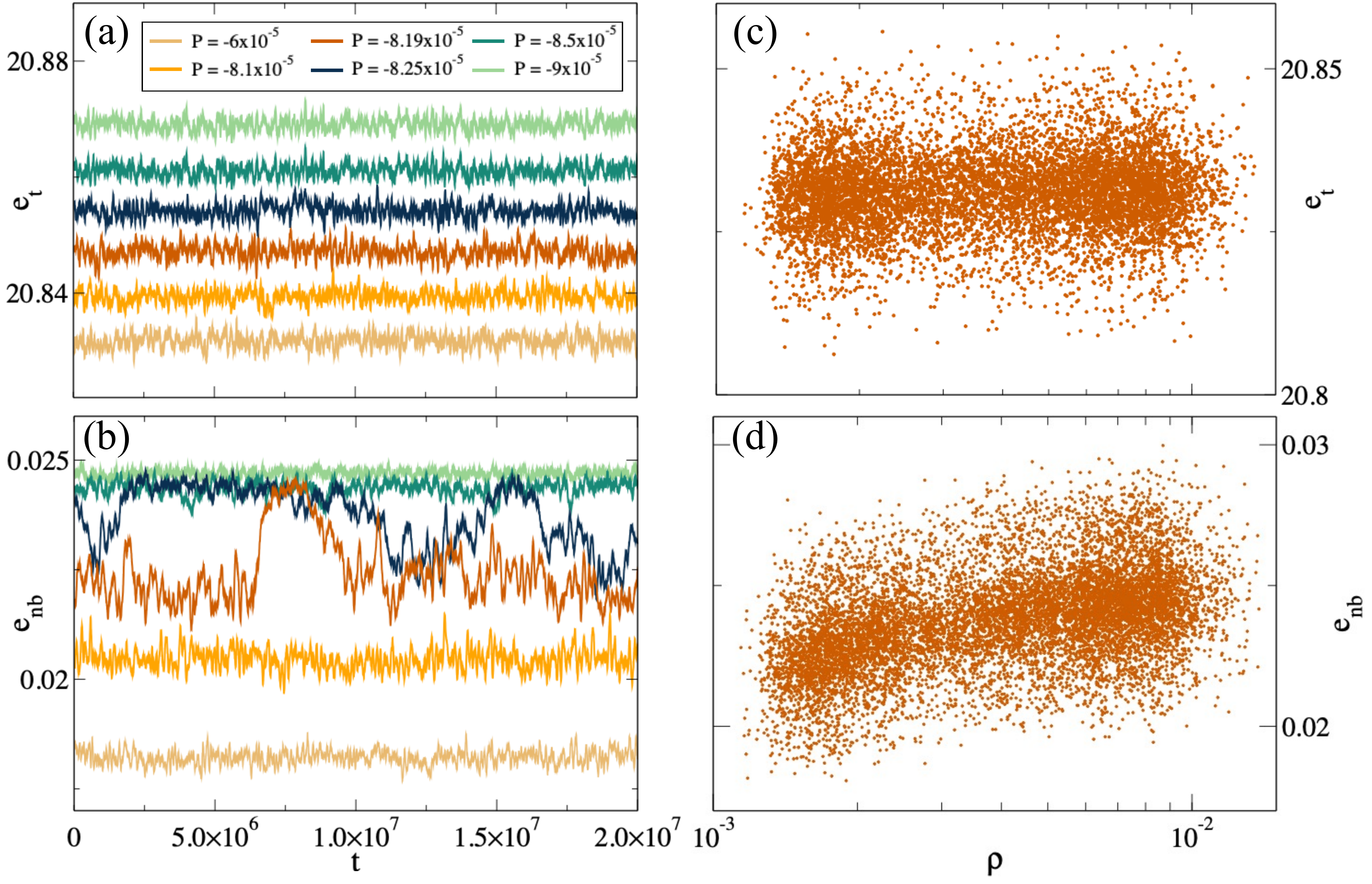}
\caption{\label{fig:ene_rho} 
Energy as a function of time for the Diam - 0.35\% network at different negative pressures: (a) total potential energy $e_t$ and (b) non-bonded particle energy $e_{nb}$. Data for $e_t$ are vertically translated by 0.05 with respect to each other to improve visualization; (c, d) scatter plots of the same quantities with respect to density $\rho$ at $P_{min}=8.19\times10^{-5}k_BT/\sigma^3$. }
\end{figure}

When we monitor the total potential energy $e_t$ as a function of time, as reported in Fig.~\ref{fig:ene_rho}(a), again for the Diam - 0.35\% network at the same value of $P$,  we find that critical fluctuations are completely absent. We thus focus on the non-bonded potential energy $e_{nb}$, shown in Fig.~\ref{fig:ene_rho}(b), which instead shows critical-like fluctuations. This correspondence is confirmed by the scatter plots, reported in  Figs.~\ref{fig:ene_rho}(c) and (d), of total energy and non-bonded energy with density, respectively. Clearly, while the former appears to be not correlated with $\rho$, the second is. This analysis allows us to focus on $e_{nb}$ as mixing variable in the order parameter $M$ that controls the criticality of the transition, as also discussed in the main text.

\newpage
\section{Scaling properties close to the transition}
\label{sec:s4}

\begin{figure}[th]
\includegraphics[width=0.9\linewidth]{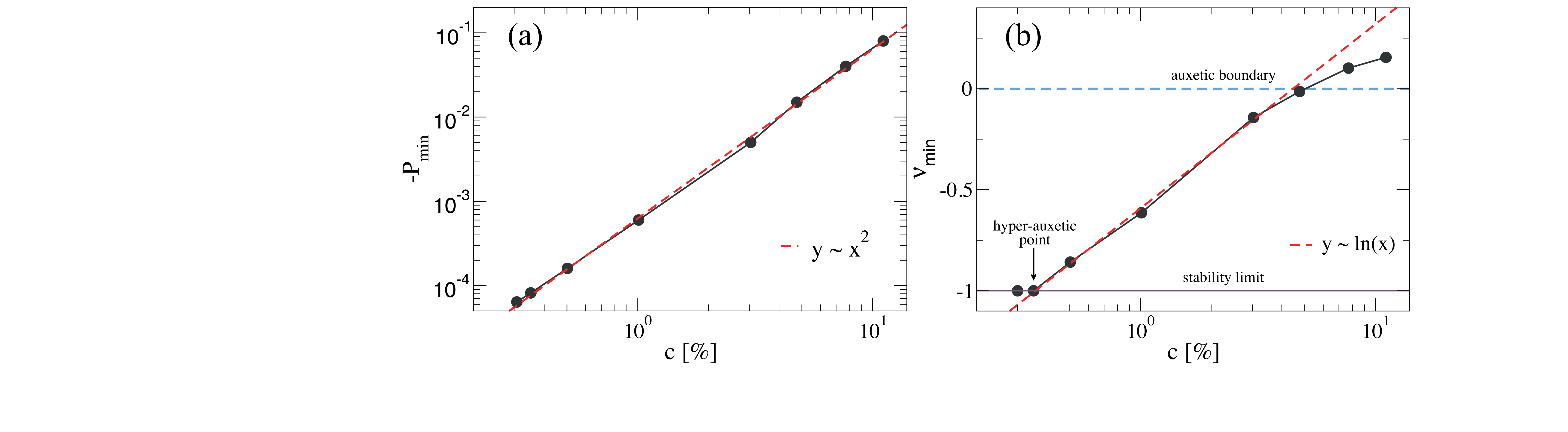}
\caption{\label{Pmin-numin} (a) Minimum pressure $P_{min}$ and (b) minimum Poisson's ratio $\nu_{min}$ for diamond networks as a function of  crosslinker concentration. Fits to the data are shown as dashed lines: in (a) a power law fit is used, while in (b) we employ a logarithmic function. The horizontal lines in (b) indicates $\nu=0$ (dashed), below which auxetic behavior starts,  and $\nu=-1$ (solid), that sets the limit of stability of the network. Note that for $c < c^*$ deviations from the logarithmic fit are observed, because the minimum Poisson's ratio always remains equal to -1. We do not include this last point in the fit.}
\end{figure}

In Fig~\ref{Pmin-numin} we report the behavior of  $P_{min}$ and $\nu_{min}$ as a function of crosslinker concentration for diamond networks. 
It is interesting that $P_{min}$ closely follows a quadratic power-law behavior for $c \lesssim 1\%$, suggesting a critical behavior that would terminate at $c=0$. Instead, $\nu_{min}$ follows a logarithmic behavior in $c$ in the auxetic region, which ends at $c^*$, below which the minimum Poisson's ratio saturates at -1, the minimum value for a mechanically stable solid. For $c < c^*$ we indeed still detected $\nu_{min}=-1$, followed by a first-order transition between the two states upon further decreasing pressure.

\bibliographystyle{naturemag}
\bibliography{bibliography}

\end{document}